\documentclass[preprint,pra,showpacs]{revtex4}

\usepackage{graphicx}
\usepackage{latexsym}

\def\<{\langle}
\def\>{\rangle}

\def\half{{\frac{1}{2}}}
\def\be{\begin{equation}}
\def\ee{\end{equation}}
\def\bea{\begin{eqnarray}}
\def\eea{\end{eqnarray}}

\def\ceff{c_{\rm{eff}}}

\begin{document}
\preprint{cond-mat} \title{Nonlocal Fermions with Local Interactions and the SYK Model}

\author{G. C. Levine}

\address{Department of Physics and Astronomy, Hofstra University,
Hempstead, NY 11549}

\date{\today}

\begin{abstract}
One of the most promising routes to non-fermi liquids and strange metals has been through SYK models \cite{Sachdev:2010um}, which necessarily involve large flavor degrees of freedom and interactions with imposed disorder. We introduce an interacting model of nonlocal spinless fermions in which large flavor and effective disorder emerge spontaneously in the extreme nonlocal limit. This model may be thought of as an expansion in a dimensionless "locality scale," $\alpha$, with the limit $\alpha \rightarrow 0$ recovering a conventional local action. For finite $\alpha$, the resulting interacting nonlocal action exhibits a large number  of low-energy degrees of freedom---proportional to $\alpha$---and the structure factor mediating the local interactions between these fermions is effectively random for large $\alpha$.  In one dimension, with finite $\alpha$, we show that interactions are marginal to the one loop level (as they are in the conventional local case), preserving a gapless phase.  At large $\alpha$ the interaction strength becomes comparable to the bandwidth and we analyze this large flavor limit in a manner similar to the diagrammatic approach to SYK. As $\alpha$ is increased we argue that the gapless phase established by conventional RG possibly crosses over to a gapless SYK phase, although the "melon" diagrams are not exclusively dominant. We speculate on how this model might arise physically from an S-matrix connecting pure excited states---rather than the vacuum---to access finite temperature interacting fermions. 
\end{abstract}


\maketitle
\section{Introduction} 


Nonlocal quantum lattice models and field theories have appeared in many subfields of physics, spanning quantum information, condensed matter physics and quantum gravity. Random models such as \cite{Sachdev:2010um} have been introduced in connection with the phenomenology of strange metals, but also appear to be important in constructing exactly solvable quantum models with gravitational duals \cite{Maldacena:2016hyu}. Nonlocal models have also appeared in the study of the thermalization hypothesis providing the rigorous basis of the canonical ensemble of statistical mechanics \cite{Magan:2015yoa}.  Attempts to understand the internal degrees of freedom of black holes, consistent with the no-cloning theorem, have led to the notion of "fast-scrambling" \cite{Hayden:2007cs,Sekino:2008he} which also can be realized with certain nonlocal models \cite{Magan:2016ojb,Swingle:2016var}.  Entanglement entropy of nonlocal bosons was first studied analytically in \cite{Li:2010dr} and numerically in \cite{Shiba:2013jja}, with nonlocality extended to fermions in  \cite{2019PhRvD.100b5017L}. There have been several analytical works on the role that nonlocal QFT plays in flat space holography \cite{Bagchi:2014iea,Kachru:2018,Pang:2014tpa}.

The motivation for the present work is threefold.  First, we introduce a model to study interactions of nonlocal fermions. This model becomes---in effect---a large flavor model and we argue that interactions may be treated by conventional renormalization group techniques.  Second, we explore the possibility of local interactions producing an effective random interaction, transforming the large flavor model into an effective SYK model.  The need for a mechanism to produce disorder in applications of SYK physics to condensed matter is self-evident, but the need for such a mechanism in applications of SYK to quantum gravity (see for instance, \cite{2019JPhA...52U4002W}) has also been emphasized.  Finally, there is the established correspondence between the ground state of fermionic nonlocal models and an excited state (with a specific energy) of the corresponding local, free fermion models \cite{2019PhRvB.100p5135J}. As such free excited states have been shown to thermalize on a sufficiently short length scale \cite{2015PhRvB..91h1110L}, nonlocal physics with interactions may provide new theoretical machinery to apply to finite temperature transport and relaxation phenomenon such as in the phenomenology of strange metals. 

To introduce this topic generally, consider the following 1-d hamiltonian for free, nonrelativistic, spinless fermions in the continuum \cite{2019PhRvD.100b5017L}:
\be
\label{NL_fermions}
H_{\rm c} =  \frac{\epsilon}{\alpha^2}\int{dx \psi^\dagger(x) \cos{(i\alpha \partial_x)} \psi(x)}
\ee
where $\partial_x$ is a spatial derivative, $\alpha$ is a dimensionful locality scale  and $\epsilon$ is an energy parameter with units of energy $\times$ length. Noting the generator of finite translations,
\begin{equation}
\label{translations}
\psi(x+\alpha) = e^{\alpha \partial_x} \psi(x) 
\end{equation}
the nonlocal kinetic operator simply expresses finite translations to adjacent points rather than a diffusion operator. The limit $\alpha \rightarrow 0$ recovers the conventional non-relativistic kinetic energy operator. Since the kinetic operator is a harmonic function, we refer to this model as "compact" nonlocality.

\begin{figure}[ht]
\includegraphics[width=11.0cm]{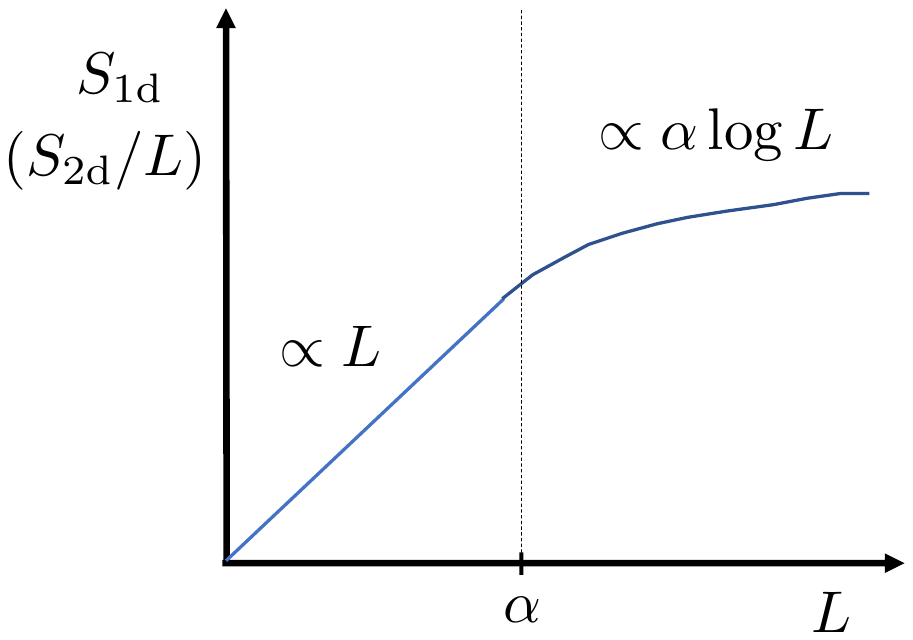}
\caption{\label{fig9} Depiction of entanglement entropy spanning local and nonlocal regimes based upon numerical calculations in \cite{2019PhRvD.100b5017L}. The 2d entropy has been rescaled by $L$. Entropy satisfies a volume law for small $L$ and a logarithmic law characteristic of gapless fermions for large $L$. The locality scale, $\alpha$, dividing nonlocal $(L<\alpha)$ and local $(L > \alpha)$ regimes is also proportional to the number of fermion flavors (the central charge) in the microscopic calculation. Thus the large $L$ (local) behavior for the entropy is proportional to $\alpha \log{L}$.  The prefactor $\alpha$ in the latter expression is also suggested "naturally" by the continuity of the graph.}
\end{figure}

The kinetic energy operator now involves derivatives of all orders and nonlocality but leads to a well-defined theory.  Placing this theory on a lattice, so that $\alpha$ is dimensionless and the fundamental degrees of freedom are finite, this hamiltonian simply leads to an energy momentum dispersion relation that is periodic in $\alpha/2\pi$. Considering an average occupancy of 1/2, it is readily seen that rather than two fermi flavors (two fermi points in 1-d), there are $\cal{O}(\alpha)$ fermi flavors.

If $\alpha/\pi$ is large and irrational, nearby energy states below the Fermi level will be drawn from widely disparate wavenumbers, maintaining the energy volume of the Fermi sea but scrambling the energy order of the translation invariant eigenfunctions. Thus the nonlocal hamiltonian has a ground state that corresponds to an excited state of the underlying local model. Quantum many body states exhibiting an entropy volume law were first realized as excited states of integrable hamiltonians \cite{Alba:2009th} and a remarkable momentum/spatial duality feature introduced in \cite{2015PhRvB..91h1110L} (see also \cite{2017PhRvL.119b0601V,Vidmar:2018rqk}). Thus there is a close connection between free nonlocal models and excited states in that the ground state of our model for a given locality parameter $\alpha$ must correspond to a particular excited state of the corresponding local model. 

In the limit $\alpha \rightarrow \infty$, the irrational values of $\alpha/\pi$ lead to random correlation functions resembling those of models with quenched disorder, although the present hamiltonian is translation invariance. This is the specific feature leading to the volume entanglement entropy law when entropy is computed for a spatial subregion smaller than $\alpha$.  Fig 1 depicts the behavior of the entanglement entropy computed in \cite{2019PhRvD.100b5017L} for regimes smaller and large than $\alpha$. To summarize, $\alpha$ appears in two roles: (1) the number of fermion flavors (specifically, int($\alpha/\pi$) = central charge c in 1-d ) (2) $\alpha$ is the locality scale designating the boundary between extensive and logarithmic entanglement entropy. 



\section{1-d nonlocal fermion model}


Consider a conventional one-dimensional noninteracting kinetic hamiltonian for spinless fermions:
\begin{equation}
\label{Ham_c}
H = \sum^{N-1}_{x,y=0}{\psi^{\dagger}_{j}S_{jk}\psi_{k}}
\end{equation}
where $j$ and $k$ are one-dimensional site indices and the operators $\psi_{j}$ ($\psi^{\dagger}_j,$) destroys (creates) fermions at site $j$ of a periodic $N$ site lattice and obey the conventional Dirac fermion algebra. $S_{jk}$ and $T_{jk}$ are the symmetric and antisymmetric lattice operators:
\begin{eqnarray}
\label{LatticeDerivatives}
S_{jk} &=& \delta_{j,k+1} + \delta_{j,k-1} \\
T_{jk} &=& \delta_{j,k+1} - \delta_{j,k-1} \nonumber
\end{eqnarray}
We now aim to write $H$ in a chiral lattice form, suited for making a nonlocal extension along the lines of equation (\ref{NL_fermions}). With the transformation to a new Dirac fermion, $d_j$,
\be
\psi_j = e^{i \frac{\pi}{2}j}d_j
\ee
the hamiltonian may be rewritten where odd and even sublattices are coupled
\begin{equation}
\label{Ham_d}
H = i\sum_{j \,{\rm (even)}}{ [d^{\dagger}_{j}(d_{j+1} - d_{j-1}) +d^\dagger_{j+1}(d_{j+2} -d_j) ]}
\end{equation}
and in matrix form,
\be
\label{matrix_Ham_d}
H = \sum_{j \in {\rm e}; k \in {\rm o}} \left(\begin{array}{c} d_{j+1} \\ d_{j} \end{array}\right)^\dagger
\left(\begin{array}{cc} 0 & i T_{jk} \\ i T_{jk} & 0 \end{array}\right) \left(\begin{array}{c} d_k \\ d_{k+1} \end{array}\right)  
\ee
where "e" and "o" refer to even and odd sublattices. $H$ may be diagonalized by introducing chiral fermions $L_j$ and $R_j$ defined as follows:
\bea
\label{chiral_ferm}
L_j &=& \frac{1}{\sqrt{2}}(d_{j+1} + d_j ) \,\,\,\,\,\,  R_j = \frac{1}{\sqrt{2}}(d_{j+1} - d_j )\\
\bar{L}_j &=& \frac{1}{\sqrt{2}}(d_{j+1}^\dagger + d_j^\dagger ) \,\,\,\,\,\,  \bar{R}_j = \frac{1}{\sqrt{2}}(d_{j}^\dagger - d_{j+1}^\dagger ) 
\eea
Note that $\bar{R}_j$ may be alternatively defined as $R^\dagger$.  Finally, $H$ may be written compactly in terms of the Pauli matrix $\sigma_z$,
\begin{equation}
\label{chiral_Ham}
H = i \sum_{j \in {\rm e}; k \in {\rm o}} { \bar{\phi}_j  T_{jk} \sigma_z \phi_k}
\end{equation}
with the definitions:
\be
\label{chiral_spinor}
\phi_j =  \left(\begin{array}{c} L_j \\ R_j \end{array}\right) \,\,\,\,
\bar{\phi}_j =  \left(\begin{array}{c} \bar{L}_j \\ \bar{R}_j \end{array}\right)
\ee

The extension to a "compact" nonlocal model as previously developed in \cite{2019PhRvD.100b5017L} is now straightforward:
\begin{equation}
\label{chiral_nonlocal_Ham}
H_\alpha = \frac{1}{\alpha} \sum_{j \in {\rm e}; k \in {\rm o}} { \bar{\phi}_j  [\sin{ (i \alpha \sigma_z  \bf{T}} ) ]_{jk}\phi_k}
\end{equation}
In the continuum limit, where $\alpha$ is dimensionful, $H$ may be written:
\begin{equation}
\label{chiral_nonlocal_Ham_continuum}
H^{\rm cont}_\alpha = \frac{1}{\alpha} \int dx \bar{\phi}(x)  \sin{(i \alpha \sigma_z  \partial_x)}\phi(x)
\end{equation}

\begin{figure}[ht]
\includegraphics[width=11.0cm]{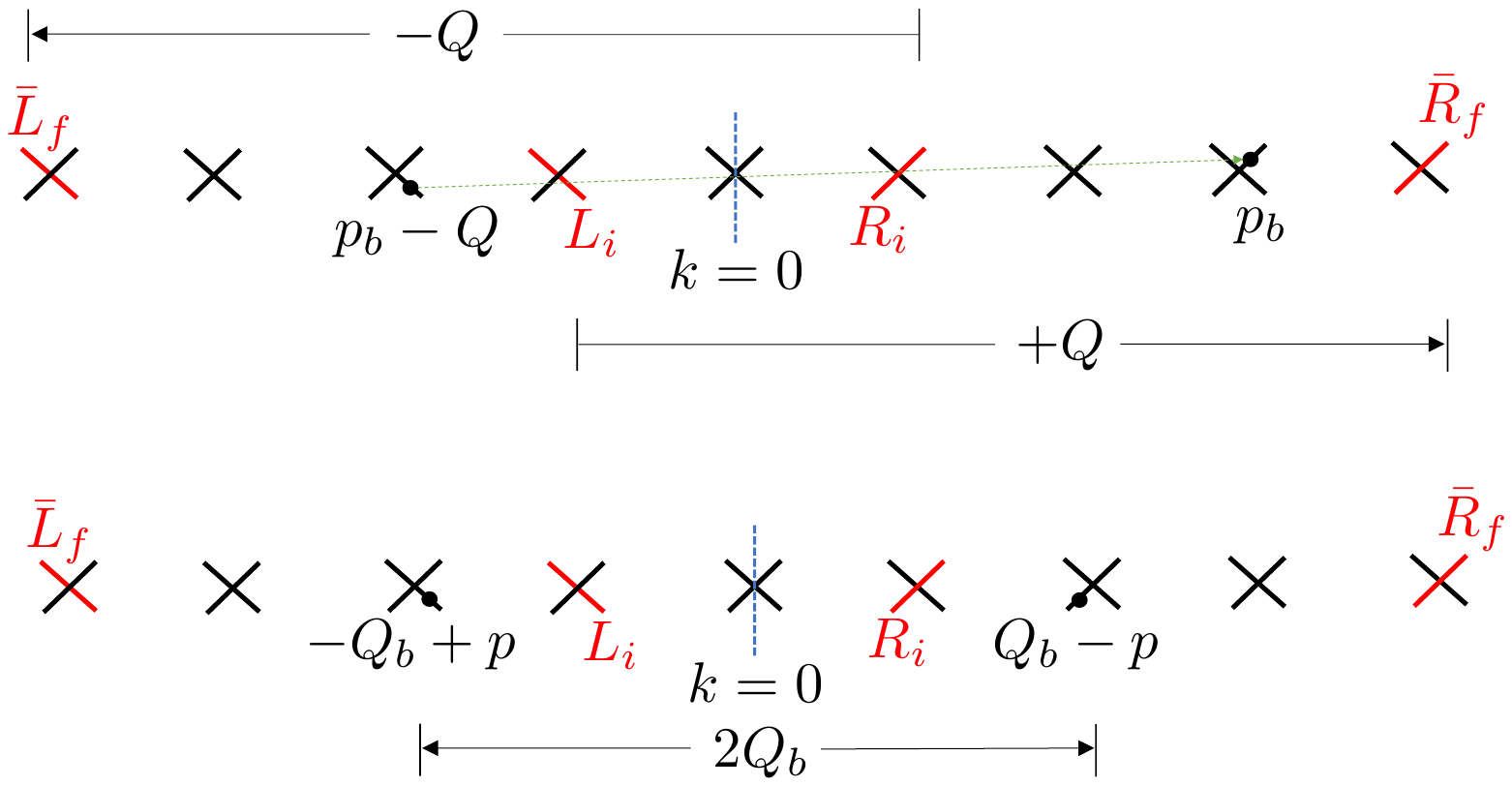}
\caption{\label{fig_multibranch} Depiction of multi-branch energy dispersion and processes corresponding to Feynman diagrams in figure (\ref{RG_diagrams}) : ZS$^\prime$ (top) and BCS (bottom).  Because of the specific form of the energy-momentum relationship in the hamiltonian $H_\alpha$ (equation \ref{H_alpha} ) the branches are not evenly spaced as shown here... This figure illustrates processes contributing to the renormalization of one particular coupling constant (for the LRLR vertex shown) of O($\alpha^2$) possible coupling constants. For the ZS$^\prime$ process (top) the nesting vector $Q$ is set by the particular choice of vertex fields $(\bar{L}_f \bar{R}_f L_i R_i)$, but the internal loop momentum $p_b$ is free to range over $\cal{O}(\alpha)$ branches, generating a singular contribution in the loop integration at each one.  For the BCS process (bottom), the nesting vector $Q_b$ is itself free to range over $\cal{O}(\alpha)$ branches generating a singular contribution at each one.}
\end{figure}

Introducing momentum space operators for the original Dirac fermions and comparable expressions for the $L_j$ and $R_j$ fermions
\be
\psi_j = \frac{1}{N}\sum_{n=-\frac{N}{2}+1}^{\frac{N}{2}}{\psi_n e^{i k_n j} } \rightarrow \psi_j = \frac{1}{2\pi}
\int_{-\pi}^{\pi}{dk \psi(k) e^{i k j}}
\ee
\be
L_j = \frac{1}{2\pi}\int_{-\pi}^\pi{dk L(k) e^{ijk}}
\ee
where $k_n = 2n\pi/N$. With $N \rightarrow \infty$, $H_\alpha$ may be written
\be
\label{H_alpha}
H_\alpha = \frac{1}{2\pi \alpha}\int_{-\pi}^\pi{dk (\bar{R}(k)R(k) - \bar{L}(k)L(k))\sin{[\alpha}\sin{k} ]}
\ee
As the locality parameter $\alpha$ grows large, the energy as a function of momentum in the Brillouin zone crosses zero $O(\alpha)$ times leading to a series of pairs of $R,L$ moving branches (depicted as in figure (\ref{fig_multibranch})). Using the nonlocal dispersion relation, the total number of branches $B$ is related to $\alpha$ by $B = 4 \lfloor \alpha/\pi \rfloor$.

Our goal is to study spinless, nonlocal fermions interacting through a local, nearest neighbor interaction.  In terms of the original Dirac fermions, this interaction is taken to be:
\begin{equation}
\label{interaction_c}
V = \frac{1}{2!2!} \frac{U}{B} \sum^{N-1}_{j=0}{\psi^{\dagger}_{j}\psi_{j} \psi^{\dagger}_{j+1}\psi_{j+1}}
\end{equation}
where the bare interaction strength has been rescaled by $B$.  Noting the $1/\alpha$ in the hamiltonian (\ref{chiral_nonlocal_Ham}) necessary to recover the local limit, it is seen that both kinetic and interacting energy scales are rescaled by $B \sim \cal{O}(\alpha)$. Such a scaling gives strong coupling in the local limit, and coupling comparable to the bandwidth in the nonlocal limit. But it should be emphasized that there is no rationale for this particular scaling of the interaction other than that it yields sensible results in both the RG analysis and SYK analysis of the model. 

In momentum space,
\begin{equation}
\label{interaction_cont}
V = \frac{1}{2!2!} \frac{U}{B}\int{\frac{dk dk^\prime dq}{(2\pi)^3} F(k,k^\prime,q) \psi^{\dagger}(k)  \psi^\dagger(k^\prime) \psi(k^\prime - q) \psi(k+q) }
\end{equation}
where the form factor, $F(k,k^\prime,q) = \half(\cos{q} - \cos{(k-k^\prime +q}))$.  The form factor may also be written in a more symmetric form:
\be
\label{form_factor}
F(\Delta_i, \Delta_f) = -\cos{\frac{\Delta_i}{2}} \cos{\frac{\Delta_f}{2}}
\ee
where $\Delta_i$ is the difference in the ingoing momenta to the vertex and $\Delta_f$ is the difference in the outgoing momenta from the vertex.

\section{RG analysis}

The low energy physics of the noninteracting nonlocal hamiltonian $H_\alpha$ is described by a collection of $B \sim \cal{O}(\alpha)$ pairs of linearly dispersing chiral modes $R(k_b), L(k_b)$ that we will label by a "branch" index $b\in(-B/2, B/2)$ index and a momentum index, $k_b$, referring to the specific branch. This series of Fermi "points" is depicted in figure 2. Although they are not evenly spaced as depicted, owing to the $\epsilon_k = \sin{(\alpha \sin{k})}$ dispersion, we will treat them as evenly spaced in much of what follows and argue that the spacing variations are irrelevant within RG. The distinct quantum fields on different branches are labeled by the momentum variables so that the notation is less cluttered. It should be stressed that, for instance, $R(k_b)$ and $R(k_{b^\prime})$ are {\sl distinct} quantum fields. This feature is supported by the inference of a central charge $c \propto \alpha$ from the asymptotic logarithmic behavior of entanglement entropy in numerical treatments of $H_\alpha$ \cite{2019PhRvD.100b5017L}.   A full treatment (for instance, within nonabelian bosonization) would include zero-mode anomalies so that these fields necessarily anticommute. We will concentrate on the renormalization of the zero momentum vertices where $(L_i, R_i) \rightarrow (\bar{L}_f, \bar{R}_f)$ are chosen to be $(L(k_{-b}=0), R(k_b=0)) \rightarrow ( \bar{L}(k_{-b^\prime}=0), \bar{R}(k_{b^\prime}=0) ) $; that is, a $\pm b$ pair of fields scattering to another $\pm b^\prime$ pair of fields.  Thus, there is a large ${\cal{O}}(\alpha^2)$ set of coupling constants, one for each $(b,b^\prime)$ choice, but they are all renormalized identically.  

\clearpage

Interactions create particle hole pairs within one branch that interact with modes of the same branch---these are Zero Sound (ZS) diagrams figure 3(a).  But there are also ZS$^\prime$ diagrams that correspond to excitations in one branch interacting with modes on another branch, as well as BCS diagrams that exchange Cooper pairs on different branches. These latter two processes are specifically depicted in figures 3 (a,b). Similar to the conventional (local) 1-d case, we will argue that ZS$^\prime$ and BCS diagrams exactly cancel---although the reason has subtleties distinct from the 1-d case---and that weak interactions are marginal to the one loop level. (As in the conventional case, ZS diagrams are nonsingular). We closely follow the formalism and notation of Shankar \cite{Shankar} in his exceptionally lucid treatment of the renormalization group for nonrelativistic fermions.

In the RG that follows, we adopt a euclidean path integral formalism for the partition function:
\be
\label{PI_RG}
Z = \int {D\bar{\psi} D\psi e^{+S}} \,\,\,\, \,\,\,\, 
S = -\int{ (\bar{\psi}\partial_\tau \psi + H_0(\bar{\psi} \psi) + V(\bar{\psi} \psi)) }
\ee
Defining fourier expansions as follows. The noninteracting action implicitly defines the greens function $G_0$ as follows:
\be
S_0 = \int \frac{dk}{2\pi} \int \frac{d\omega}{2\pi} \bar{\psi}(k \omega) \psi (k \omega) (i\omega - \epsilon_k + \mu) = 
\int \frac{dk}{2\pi} \int \frac{d\omega}{2\pi} \bar{\psi}(k \omega) G_0^{-1} \psi (k \omega)
\ee
where $\epsilon_k$ and $\mu$ are the energy dispersion and chemical potential for $H_0$.

\begin{figure}[ht]
\includegraphics[width=11.0cm]{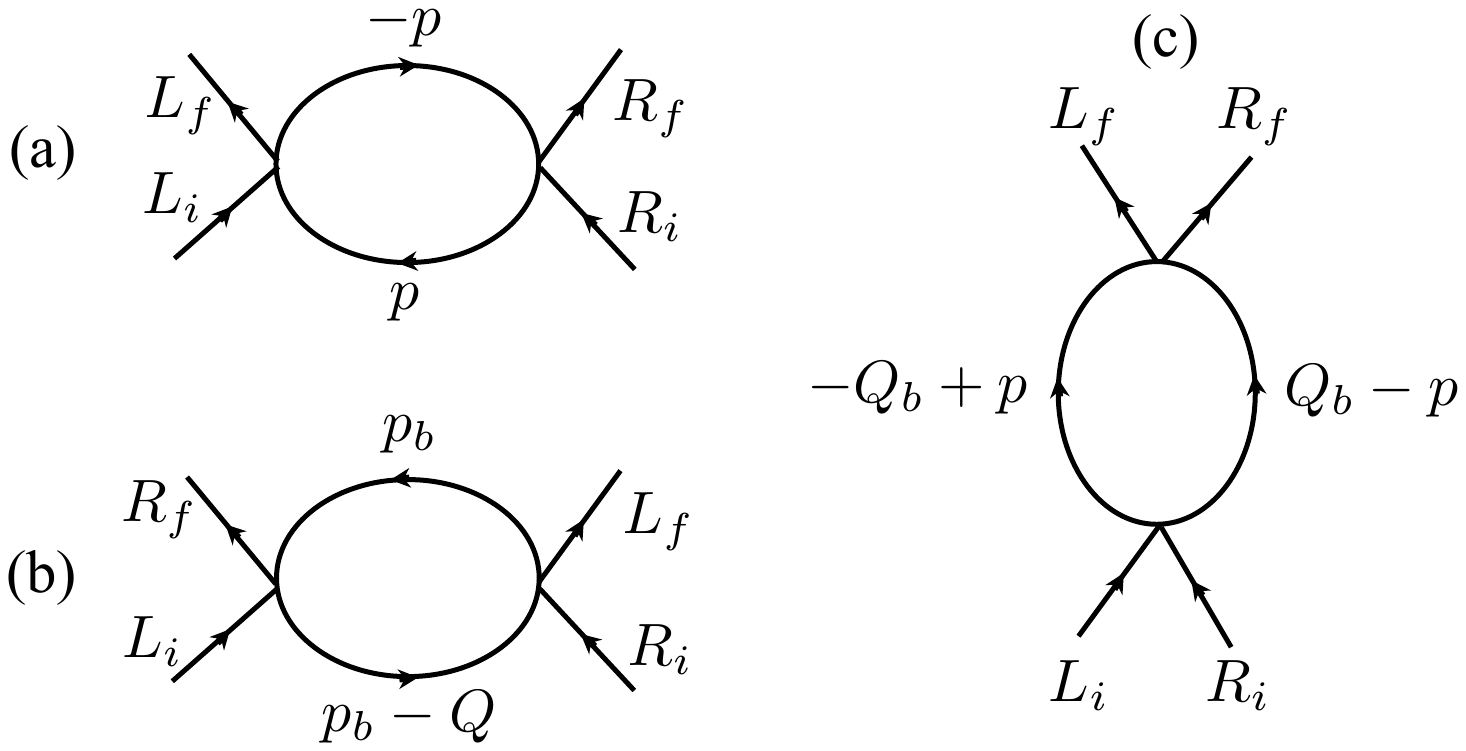}
\caption{\label{RG_diagrams} (a) Feynman diagram for Zero Sound (ZS), (b) another Zero Sound channel (ZS') and (c) Pairing (BCS). As in the local case, the ZS diagram is finite and does not contribute to the RG analysis.}
\end{figure}

{\bf ZS$^\prime$ diagrams}.  Even though the RG analysis is completely standard, it is useful to show sample contractions for one diagram in detail because of the inherent complication of multiple branches.  For simplicity, frequency variables have been suppressed. Expanding the interaction in equation (\ref{PI_RG})  to second order, we make assignments of $\psi$ and $\bar{\psi}$ to four external lines following figure 3(b). The form factors $F(k,k^\prime,q)$ are taken to be unity at this order.

\begin{eqnarray}
\label{second_order}
\int{dk_1 dk_2 dq_1} \int{dk_3 dk_4 dq_2}  F(k_1 k_2 q_1) F(k_3 k_4 q_2) \times \half ( \frac{U}{2!2!})^2 \times \nonumber \\
\underbrace{\bar{\psi}(k_1)} \underbrace{\bar{\psi}(k_2)} \underbrace{\psi(k_2 - q_1)} 
\underbrace{\psi(k_1 + q_1)} \,\,\,\, 
\bar{\psi}(k_3) \bar{\psi}(k_4) \psi(k_4 - q_2) \psi(k_3 + q_2)
\end{eqnarray}
\be
\bar{L}_f \,\,\,\,  \bar{R}(p_b) \,\,  L(p_b - Q) \,\,\,\,  \,\,\,\,  R_i  \,\,\,\,   \,\,\,\,  \,\,\,\,  \,\,\,\,  \bar{R}_f  \,\,\,\, \bar{L}(p_b-Q) 
\,\,\,\, R(p_b) \,\,\,\,  L_i.  \,\,\,\,  \,\,\,\, 
\ee
As usual, the multiplicity in the choice of external lines cancels the $(1/2!2!)^2$ factor, and the two possible assignments of in/out modes to vertices cancels the $1/2$ factor in the exponential expansion.  Note that $Q$, the analog of a "nesting" vector in a conventional Brillouin zone, is fixed by the choice of external fields whereas $b$, the index of the internal loop momentum, is free to range over all branches. Adopting a canonical order for the external lines of the vertex based upon equation \ref{interaction_cont}, the ZS$^\prime$ diagram may be expressed symbolically, ZS$^\prime$ = $\bar{L} \bar{R} L R (-GG)$.  The correction to the coupling constant $U$ from one loop integration is then:
\be
\Delta U_{\rm ZS^\prime} = -(\frac{U}{B})^2 \int \frac{dp_b}{2\pi} \int \frac{d\omega}{2\pi} 
\frac{1}{i\omega - \epsilon_{p_b}} \frac{1}{i\omega - \epsilon_{p_b - Q}}
= +(\frac{U}{B})^2 \int \frac{dp_b}{2\pi} \int \frac{d\omega}{2\pi} 
\frac{1}{\epsilon_{p_b} - \epsilon_{p_b - Q}}
\ee 
For a given branch index, $b$, and nesting vector, $Q$, the modes to be integrated out may lie within any one of four cutoff regions as depicted in figure \ref{ZS_cutoff_regions}.  The flow of the coupling constant corresponding to integrating out the designated modes over an infinitesimal region $\log{s} = dl$ may be written:
\be
\Delta U_{\rm ZS^\prime} = 4 \times B U^2 \frac{1}{2\pi} \int_{\Lambda/s}^{\Lambda}{\frac{dk}{2k} } = B \frac{U^2}{\pi} \Delta l
\ee
where we have included the range of $p_b$ over $B$ branches. Since the bare vertex is proportional to $U/B$, the flow of the coupling constant from ZS$^\prime$ diagrams is
\be
\frac{dU}{dl} = \frac{U^2}{\pi}
\ee

\begin{figure}[ht]
\includegraphics[width=11.0cm]{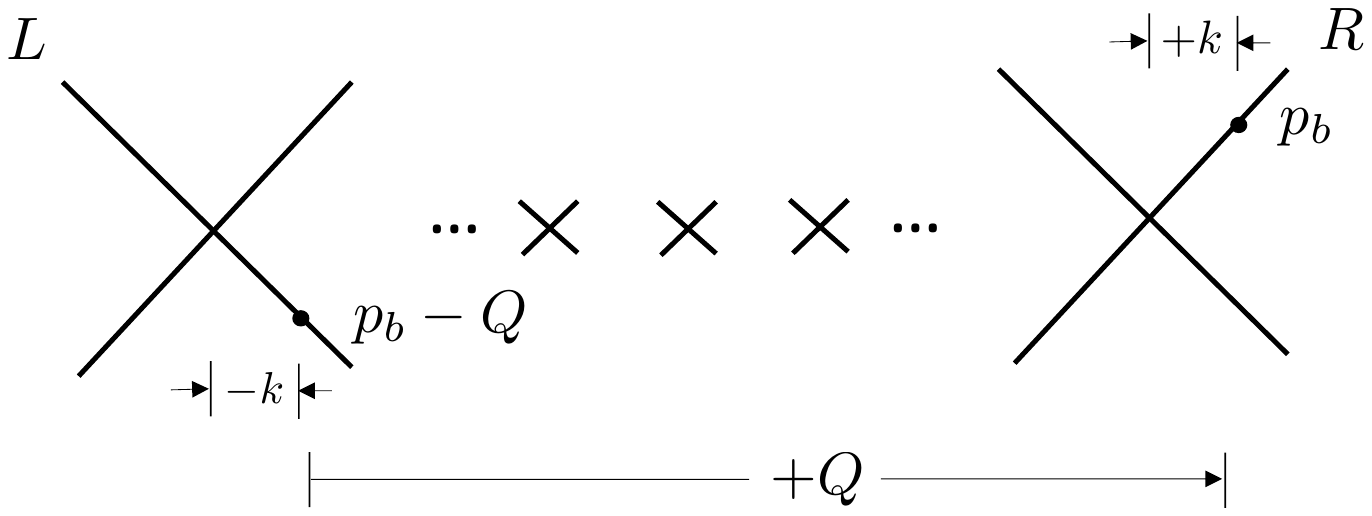}
\caption{\label{ZS_loop_momenta} Depiction of ZS$^\prime$ loop momenta. The nesting vector, $Q$ is fixed by the momenta of external lines. $p_b \in [-\pi,\pi]$ is free to range over branches $b$; $k$ is a momentum variable relative to the particular branch $p_b$ lies on. As in the conventional (local) spinless fermion case, the energy of a mode $\epsilon_k = k$.}
\end{figure}

\begin{figure}[ht]
\includegraphics[width=11.0cm]{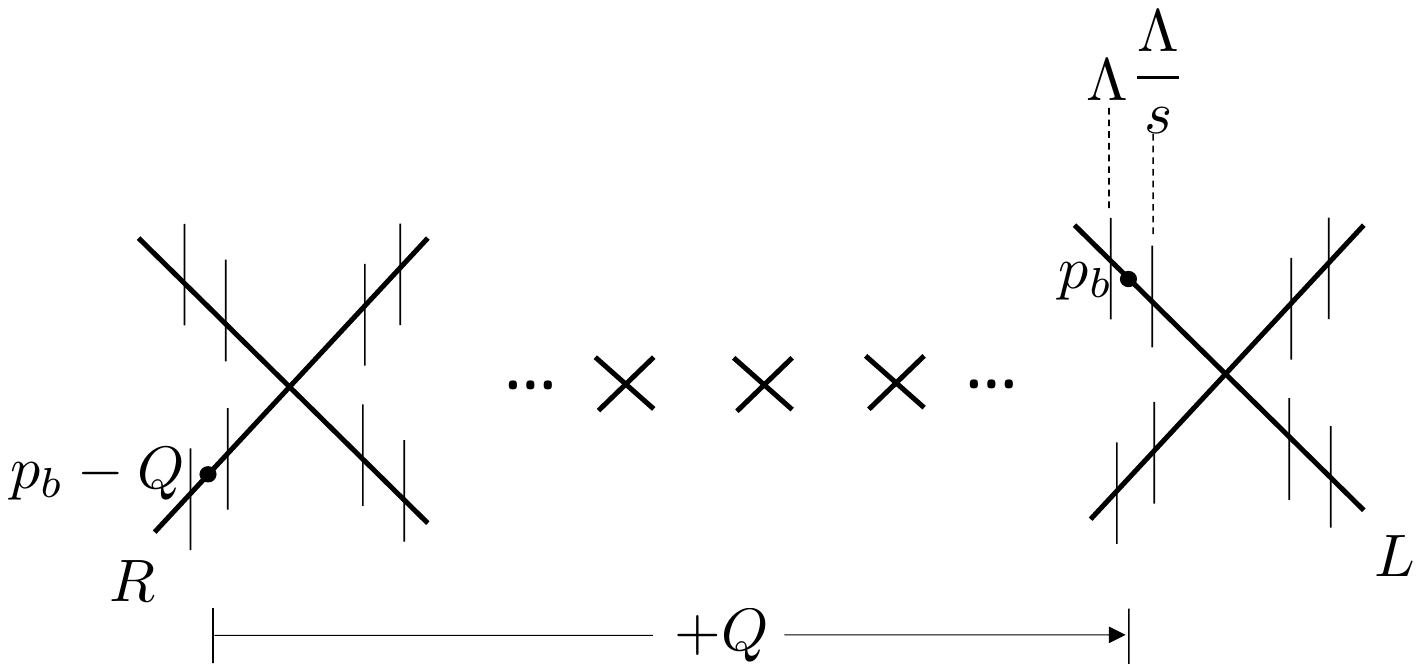}
\caption{\label{ZS_cutoff_regions} Depiction of ZS$^\prime$ loop momenta indicating four cutoff regions. $p_b$ is assigned to one of these four regions and modes between $\Lambda$ and $\Lambda/s$ are integrated out.  In this instance $p_b > 0$ is located on $L$ mode and $p_b - Q$ on an $R$ mode.}
\end{figure}

\begin{figure}[ht]
\includegraphics[width=11.0cm]{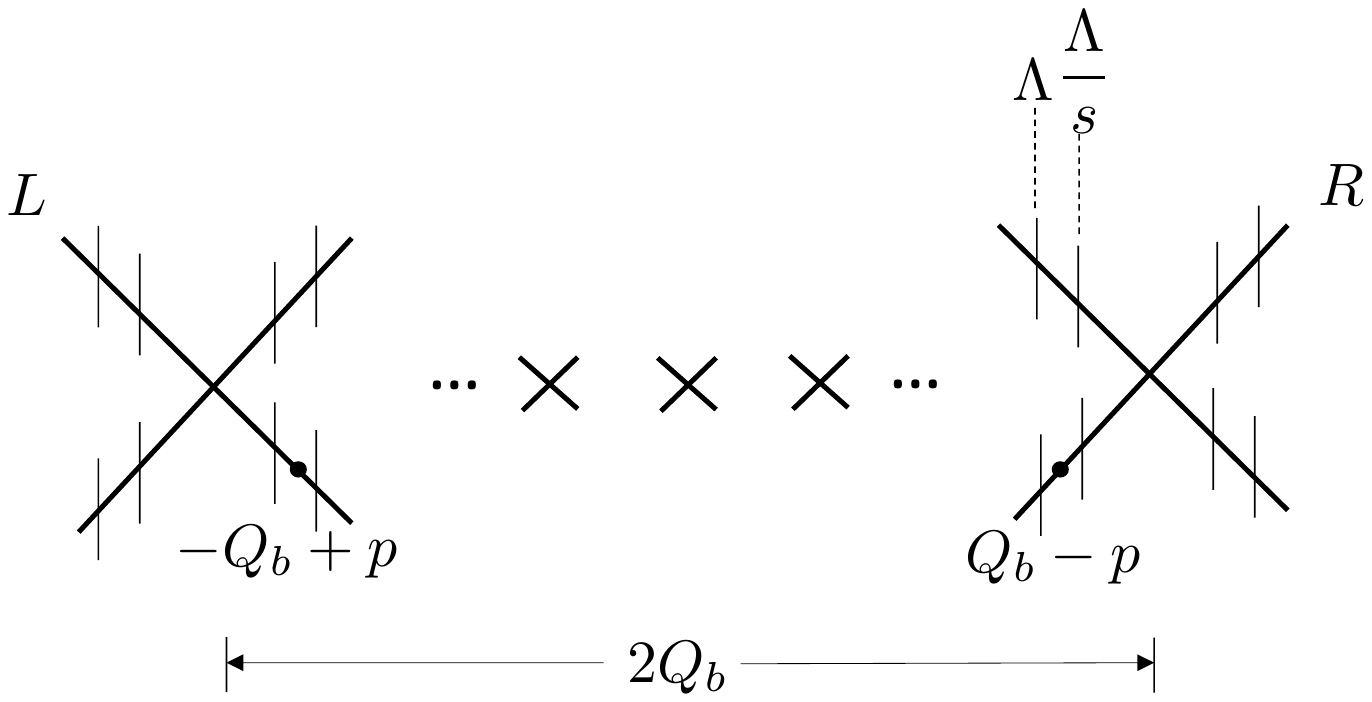}
\caption{\label{BCS_cutoff_regions} Depiction of BCS diagram momenta. In the BCS case, the nesting vector $Q_b$ is free to range over branches $b$. In contrast to the ZS$^\prime$ diagrams, there are eight cutoff regions. $Q_b-p$ is assigned to one of these eight regions and modes between $\Lambda$ and $\Lambda/s$ are integrated out.}
\end{figure}

{\bf BCS diagrams}. Examining the contractions as in equation (\ref{second_order}) there are eight ways of choosing external lines and two ways of choosing internal contractions for each leaving a net factor of $1/2$ (the diagrammatic loop "symmetry" factor) from the $\half (\frac{1}{2!2!})^2$ factor in the expansion to second order.

\be
\Delta U_{\rm BCS} = -\half (\frac{U}{B})^2 \int \frac{dp}{2\pi} \int \frac{d\omega}{2\pi} 
\frac{1}{-i\omega + p} \frac{1}{i\omega + p}
\ee 

Refer to figure showing nesting, momentum integration regions. Now including the range of the nesting vector $Q_b$ over $B$ branches, the flow of the coupling constant from BCS diagrams  corresponding to integrating out the designated modes over an infinitesimal region $\log{s} = dl$ may be written:
\be
\Delta U_{\rm BCS} = -8 \times \half B U^2 \frac{1}{2\pi} \int_{\Lambda/s}^{\Lambda}{\frac{dp}{2p} } = -B\frac{U^2}{\pi} \Delta l
\ee
Again, since the bare vertex is proportional to $U/B$, the flow of the coupling constant from BCS diagrams is
\be
\frac{dU}{dl} = -\frac{U^2}{\pi}
\ee

The two contributions cancel and thus, to the one loop level, interactions are exactly marginal.

\section{Relationship to SYK4}
One might consider whether this gapless critical state persists as the locality scale $\alpha$ is increased arbitrarily.  At some point, performing momentum shell integrations about each fermi point and ignoring the scattering between fermi points (analogous to Umklapp) becomes unphysical. Since the number of branches, $B$, has becomes large---in some sense, of measure equivalent to the momentum integrations in a single branch---we argue that the sensible model in this limit is a collection of $B$ independent fermions, lying at a sequence of fermi points, that interact through the local interaction (\ref{interaction_c}). Considering the form factor (\ref{form_factor}), summing the momenta of a high order Feynman diagram is expected to produce rapidly oscillating phases in that the momenta of the fermi points are not rationally related (for arbitrary large $\alpha$).  Thus the momentum sums produce an effect resembling disorder averaging \cite{Sachdev:2010um}. 

To this end, we rewrite the four-fermion interaction (\ref{interaction_c}) truncating the degrees of freedom to the   fermions lying at the sequence of $B$ fermi points, as described above. The effective model for the interactions of $B$ fermions flavors follows (where we have dropped the combinatoric factor in (\ref{interaction_c}) for simplicity):
\be
H = \frac{U}{B}\sum_{lkq}^B{F(l,k,q) \psi^{\dagger}_{k}\psi_{k-q} \psi^{\dagger}_{l}\psi_{l+q}}
\ee
To be clear, the $B$ flavors represent fermions with a specific (conserved) momentum, thus the sum is intended represent $B$ points in a $[-\pi,\pi]$ Brillouin zone corresponding to the zeroes of the nonlocal dispersion relation. Compared with SYK4 this interaction has the added constraint of momentum conservation and the form factor $F(l,k,q)$ replaces the interaction weight in SYK that is chosen from a random ensemble. Going over to Grassman variables the action is now:
\be
\label{multibranch_action}
S = -\int{ (\sum_{j}^B\bar{\psi}_j(\partial_\tau-\mu) \psi_j + H(\bar{\psi}_j \psi_j ))d\tau }
\ee
and the free propagator follows:
\be
G_0^{j}(\tau) = -\langle  \psi_j(\tau) \bar{\psi}_j(0) \rangle 
\ee
Following the general scheme of SYK, we compute the second order correction to the propagator corresponding to the melon diagram shown in figure 7(a):

\begin{eqnarray}
 \langle  \psi_i(\tau) \bar{\psi}_i(0) \rangle  & = & G_0^i(\tau) + \\
 & + & \half \frac{U^2}{B^2}  \sum_{p,q} F_1^2(p,q)\int_0^{\tau_2}{d \tau_1 } \int_{\tau_1}^{\tau_2}{d \tau_2}
  G_0^i(\tau-\tau_2) G_0^i(\tau_1) \times \\
& \times & G_0^{i-q}(\tau_2-\tau_1) G_0^{i-p}(\tau_2-\tau_1) G_0^{i-q-p}(\tau_1-\tau_2) 
\end{eqnarray}
The propagators $G_0^i(\tau)$ are all dispersion-less and therefore identical; that is, the only importance of the momentum superscript is to identify the reduction in phase space of the diagram due to momentum conservation. The form factor for both vertices evaluates to:
\be
F_1(p,q) = \half(\cos{p} - \cos{q}) = -F_2(p,q)
\ee
To perform the momentum sums, we approximate the distribution of fermi points in the BZ to be $B$ evenly spaced points, $p_j \approx 2 \pi j/B$. Thus, for instance,
\be
\sum_{j=-B/2}^{B/2}{\sin^2{p_j}} \approx \frac{B}{2\pi} \int_{-\pi}^\pi {dp \sin^2p} = B \overline{\sin^2 p} = \frac{B}{2}
\ee
Noting that $\overline{\sin{p}\cos{p}} = 0$, the double momentum sums produce a $B^2$ factor canceling reciprocal $B^2$ factor from the vertex. 
\begin{eqnarray}
 G(\tau)  & = & G_0^i(\tau) + \\
 & + & \frac{1}{8} U^2  \int_0^{\tau_2}{d \tau_1 } \int_{\tau_1}^{\tau_2}{d \tau_2}
  G_0(\tau-\tau_2) G_0(\tau_1) G_0^2(\tau_2-\tau_1) G_0(\tau_1-\tau_2) 
\end{eqnarray}
The success of SYK relies critically upon the suppression of higher order loop diagrams (such as figure 7(b)) through disorder averaging.  In the melon diagram computed above, loop momenta appear in even powers of harmonic functions and thus have a nonzero average.  To assess the suppression arising from the averaging over harmonic functions in the structure factor, we examined the higher loop diagrams in figure 7(b). In the expression below, we have only reproduced the sums over structure factors for the diagram given with external momentum $i$ and expressed the internal propagators symbolically.
\begin{eqnarray}
\label{syk_expansion}
 \frac{1}{B^4}  \sum_{pqrk}{ G\ldots G \times \half(\cos{p} - \cos{q}) \half (\cos{(q-r)} - \cos{p+r} ) } \times \\  \half (\cos{(i-p-k)} + \cos{r} )  \half (\cos{(i-q-k+r)} + \cos{r} ) 
\end{eqnarray}
Of the 16 terms, only 2 survive the momentum averaging. For a term to survive, all momentum labels must appear an even number of times in the product of harmonic functions. for instance, the first term in expression (\ref{syk_expansion}) $  \cos{p} \cos{(q-r)}\cos{(i-p-k)}\cos{(i-q-k+r)}  $, has a nonzero average, but  $  \cos{p} \cos{(p-r)}\cos{(i-p-k)}\cos{(i-q-k+r)}  $  vanishes summed over free internal momenta. As the number of internal loops grow, the relative number of factors containing only even pairings is expected to vanish if the internal momenta appear randomly within the harmonic functions.  

This assertion leads to an estimate of the contribution of diagrams containing internal loops.  The melon diagram nominally contains two free momenta, and each additional internal loop contains two additional momenta. Such a diagram containing $N$ vertices will contribute the product of $N$ harmonic function factors coming from the form factor, each containing some combination of $N$ free internal momenta. If we assume that in a high order diagram, the combination of free momenta within a given harmonic function is essentially random, the probability of any given momentum variable appearing an even number of times is 1/2.  Thus an ${\cal{O}}(U^N)$ diagram that is should be suppressed by a factor of $(1/2)^N$. 

In summary, internal bubble diagrams which include corrections to the two particle propagator {\sl do not vanish}, as they do under disorder averaging in the SYK model. This incomplete disorder averaging may be traced to the additional constraint of momentum conservation in the action (\ref{multibranch_action}).  High order in these internal bubble diagrams are exponentially suppressed and we thus regard the sum of melon diagrams as an uncontrolled approximation. 


\begin{figure}[ht]
\includegraphics[width=11.0cm]{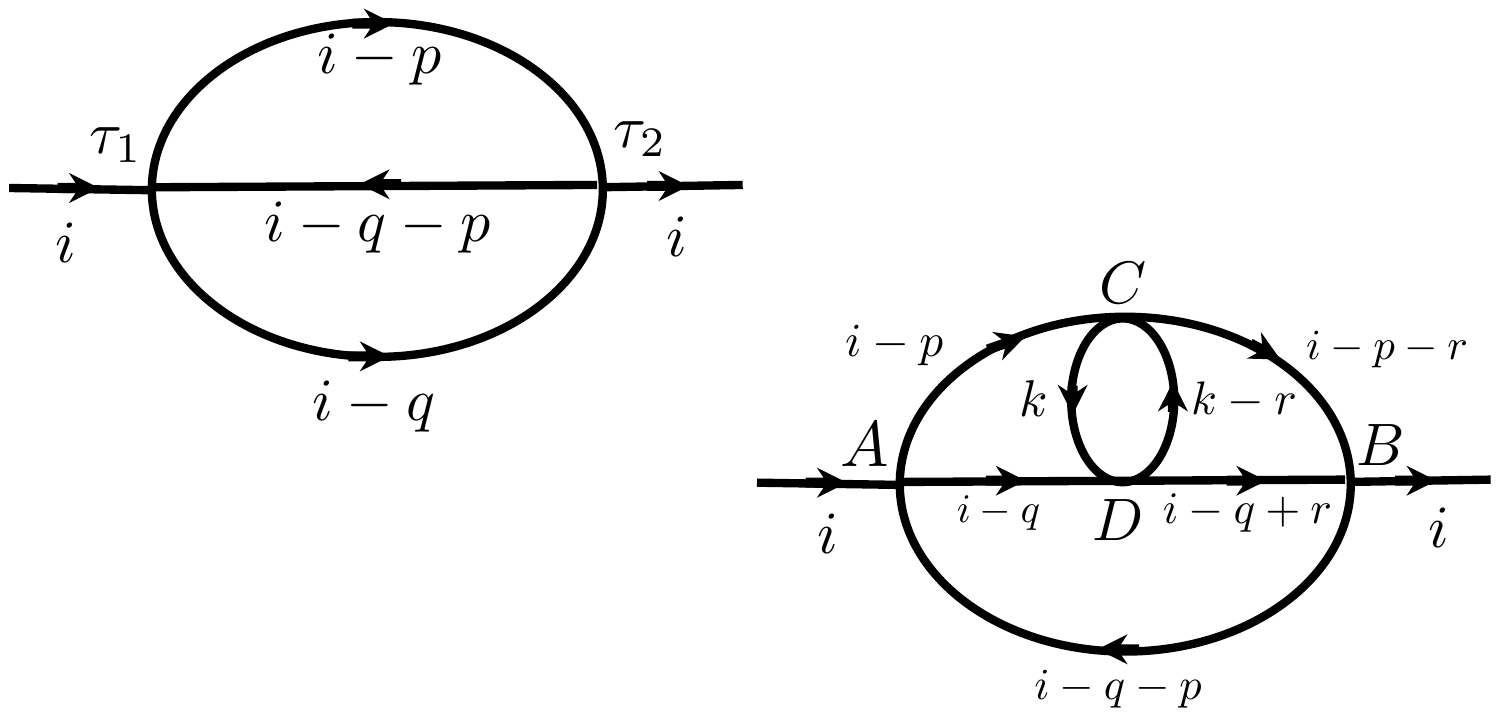}
\caption{\label{SYK_diagrams} SYK melon diagram (a) and melon diagram with internal loop (b).}
\end{figure}

\section{Discussion and Further Directions}
In this manuscript we study local density-density interactions in a 1-d model of nonlocal fermions based upon a hamiltonian (\ref{chiral_nonlocal_Ham}) and (\ref{chiral_nonlocal_Ham_continuum}). The free version of this model exhibit a crossover between volume and anomalous (logarithmic) area law behavior at scales larger than the locality scale, $\alpha$.  Owing to the periodic nature of the hamiltonian $\alpha$ is also a measure of the number of fundamental degrees of freedom in the theory---the effective central charge, $\ceff$. At weak coupling, we find that interactions are exactly marginal and the system remains gapless.  As $\alpha \rightarrow \infty$ we argue that the interaction becomes effectively random due to the rapidly fluctuating phase of the interaction structure factor written in momentum space.  Due to momentum conservation, this quasi-randomness is not as complete as the quenched disorder in the SYK model and the melon diagrams are not exclusively dominant.  However, we argue that higher order non-melon diagrams are suppressed and that an SYK phase may be stabilized.  

We now speculate about the possible physical origin of a nonlocal model such as (\ref{chiral_nonlocal_Ham}). In effect, it might follow from reinterpreting the path integral constructed for (\ref{chiral_nonlocal_Ham}) as representing amplitudes connecting a pure excited state---rather than the vacuum. It has been shown that typical excited eigenstates in a model of free fermions manifest {\sl thermal} behavior on a sufficiently short spatial scale (call this scale $\alpha$). Specifically, the reduced density matrix of a single typical excited state matches the Fermi-Dirac distribution with a temperature reflecting the average single particle energy of the excited state. Informally, this result suggests that a calculation of finite temperature interacting fermions might be approached from the S-matrix of a pure {\sl excited} state rather than a thermal ensemble. 

To explore this approach, one might regard the density matrix composed from hamiltonian (\ref{chiral_nonlocal_Ham}) as a projection operator onto an excited state of the underlying local model.
\be
\rho_0(\alpha) =  \lim\limits_{\beta \to \infty} e^{-\beta H_\alpha} 
\ee
and consider time evolution given by the {\sl local} hamiltonian $H_0 + V$ (defined in equations (\ref{chiral_nonlocal_Ham}) and (\ref{interaction_c})). Also note that we take the fictitious temperature $\beta \rightarrow \infty$ to select a single excited state. There are various ways of including interactions of a system prepared in an initial nonequilibrium density matrix (e.g. Schwinger-Keldysh closed time-loop \cite{Kamenev}, Kadanoff-Baym \cite{KadanoffBaym}, and, most recently, \cite{2019PhRvB..99e4306C}) In the closed time loop approach, the Keldysh component of the Greens function will carry the single particle distribution function formed from the initial density matrix, $\rho_0(\alpha)$; thus, this distribution function describes multiple 1-d Fermi surfaces such as depicted in figure 2. Treating the fermion fields at these fermi points as independent fields and taking the interaction energy scale as much larger than the kinetic scale of the local model ($V>>H_0$) approximately realizes the large flavor model studied in section IV.  In principle, one might be able to study transport properties---for instance, the decay of a current---generated "on top" of a zero current excited state. Specifically, the Keldysh Dyson equation (or quantum kinetic equation) governs the relaxation of the distribution function to determine transport properties. Unfortunately, the procedure sketched here may not work for the "typical" excited states that produce thermal RDMs as these excited (pure) states may not necessarily be written in terms of exponentials of single particle operators and thus obey Wick's theorem.  However, the novel approach to time evolution of athermal inital density matrices, recently proposed in \cite{2019PhRvB..99e4306C}, may be useful.

The nonlocal model may be a route to studying finite temperature fermion systems with strong interactions. The potential benefit of such an approach is that physical temperature---appearing for instance in a transport calculation---is encoded (through the energy of the excited state) in a parameter that represents the number of fundamental degrees of freedom, the effective central charge.  Just as central charge appears in thermodynamic and transport quantities conventionally, the physical temperature would appear through the parameter $\alpha$ in a zero temperature nonlocal model.

We should point out that extensions to higher dimensional lattices might be accomplished by considering a generalization of (\ref{NL_fermions}) in which a scalar diffusion operator is constructed that respects the underlying lattice orthogonality. One possible extension in 2-d might be found by following the scheme of the Dirac equation, 
\be
\label{2d_NL_fermions}
H_{\rm 2d} =  \frac{\epsilon}{\alpha^2}\int{d^2x \psi^\dagger(x) \cos{(i\alpha \sigma^\mu \partial_\mu)} \psi(x)}
\ee
where $\{ \sigma^\mu \} $ are two anticommuting Pauli matrices \cite{BB}. Similarly, this model becomes a large flavor model (with additional flavors resulting from the $\sigma$ matrices) and interactions may also be studied by conventional RG techniques. Perturbation theory and RG in both 1-d and higher dimensional models would benefit from support by numerical investigation. Unfortunately, it is likely that the nonzero locality length (and corresponding volume entropy law) would lead to a rapid increase in the required hilbert space dimension ($O(\exp{L^d})$) even in a 1-d DMRG numerical scheme and, therefore, direct diagonalization may be the only possible approach. 

Lastly, we remark that this large flavor model resembles the relativistic Gross-Neveu (GN) model. The essential difference is that the scalar interaction term in GN leads to the ZS' diagrams without the BCS diagrams.  In condensed matter parlance, the charge density wave symmetry breaking dominates superconductivity (when the ZS' channel dominate the BCS channel). The charge density wave phase is essentially equivalent to the UV asymptotic freedom associated with the GN model.




\end{document}